\documentclass{Interspeech2024}
\usepackage{hyperref}
\usepackage{multirow}
\usepackage{tabularx}
\usepackage{url}

\interspeechcameraready

\title{VoxBlink2: A 100K+ Speaker Recognition Corpus and the Open-Set Speaker-Identification Benchmark}
\name[affiliation={1,3}]{Yuke}{Lin}
\name[affiliation={1,3}]{Ming}{Cheng}
\name[affiliation={2}]{Fulin}{Zhang}
\name[affiliation={2}]{Yingying}{Gao}
\name[affiliation={2}]{Shilei}{Zhang$^\ast$}
\name[affiliation={1,3}]{Ming}{Li$^\ast$}

\address{
$^1$School of Computer Science, Wuhan University, Wuhan, China \\
$^2$China Mobile Research Institute, Beijing, China \\
$^3$Suzhou Municipal Key Laboratory of Mutimodal Intelligent Systems, \\ Duke Kunshan University, Kunshan, China 
}


\email{ming.li@whu.edu.cn}
\begin{document}

\maketitle

\renewcommand{\thefootnote}{\fnsymbol{footnote}}
\footnotetext{$^\ast$ Corresponding authors: Shilei Zhang, Ming Li. }
\setcounter{footnote}{0}
\renewcommand{\thefootnote}{\arabic{footnote}}
\begin{abstract}
    In this paper, we provide a large audio-visual speaker recognition dataset, VoxBlink2, which includes approximately 10M utterances with videos from 110K+ speakers in the wild. This dataset represents a significant expansion over the VoxBlink dataset, encompassing a broader diversity of speakers and scenarios by the grace of an optimized data collection pipeline. Afterward, we explore the impact of training strategies, data scale, and model complexity on speaker verification and finally establish a new single-model state-of-the-art EER at 0.170\% and minDCF at 0.006\% on the VoxCeleb1-O test set. Such remarkable results motivate us to explore speaker recognition from a new challenging perspective. We raise the Open-Set Speaker-Identification task, which is designed to either match a probe utterance with a known gallery speaker or categorize it as an unknown query. Associated with this task, we design concrete benchmark and evaluation protocols. The data and model resources  can be found in \url{http://voxblink2.github.io}.
    
\end{abstract}

\section{Introduction}

Speaker recognition has been widely studied over the past decades, resulting in tremendous performance improvements. Although numerous efforts have been focused on various model structures\cite{he2016deep,thienpondt2023ecapa2,wang23ha_interspeech,desplanques20_interspeech,zhang22h_interspeech} under complex application scenarios\cite{farfield_xiaoyi,qin22_interspeech,han21c_interspeech,xia2019cross}, there is still a significant gap towards the requirement of commercial applications. Given the prevailing trend of large models across different domains, we anticipate that large scaled datasets and models performance dramatically.

The variability and quality of data play essential roles in the development of robust speaker recognition systems. The VoxCeleb\cite{Nagrani17,chung18b_interspeech} is currently the most popular database for speaker recognition. Nevertheless, when compared to datasets for facial recognition, the disparity in size is profound, spanning two orders of magnitude. The VTL \cite{torgashov2023id} includes a huge amount of utterances from 100K+ speakers, but only a small version\cite{yakovlev23_interspeech} of pure audio with 5,040 speakers is publicly available. The VoxBlink dataset\cite{lin2023voxblink} introduces a novel, highly scalable data-mining method for data collection. However, it did not bring about a qualitative improvement in data volume. Hence, we have refined the data collection pipeline of VoxBlink and expanded the scope of retrieval, thereby aggregating a much bigger scale audio-visual dataset for many possible applications. Moreover, we discover that further increasing the size of data and the complexity of models can achieve \textbf{state-of-the-art} results, which drives us to explore a more challenging scenario.

While the verification task requires only a 1: 1 comparison, the identification problem requires 1: N comparisons – of a probe sample with templates from identities in a gallery. Depending on the gallery size N, finding the correct answer can be much more challenging than simply performing a correct 1: 1 comparison. Moreover, the unseen individual in the probe set should not be linked to the gallery identities (like the entrance guard, etc.) in many scenarios. Therefore, we raise the Open-Set Speaker-Identification (OSSI) benchmark based on the VoxBlink-clean set, incorporating over ten thousand speakers in evaluation for the first time. We adopt the Detection and Identification Rate at the False Alarm Rate (DIR@FAR)\cite{jain2011handbook}, which works under an m: n open-set protocol, with the ability to identify enrolled speakers and reject unseen identities. In general, our contribution can be summarized as follows: 

\begin{itemize}
\item We optimize the data collecting pipeline and release a large audio-visual speaker recognition dataset at the scale of 100K+ speakers.
\item We explore different training strategies with different model sizes and data volumes to show the trend, and achieve the state-of-the-art system performance on the Vox1 test set.
\item We propose a new Open-Set Speaker-Identification benchmark as well as evaluation protocols and baselines.
\end{itemize}

\begin{figure}
\vspace{2.5em}
  \includegraphics[width=0.47\textwidth]{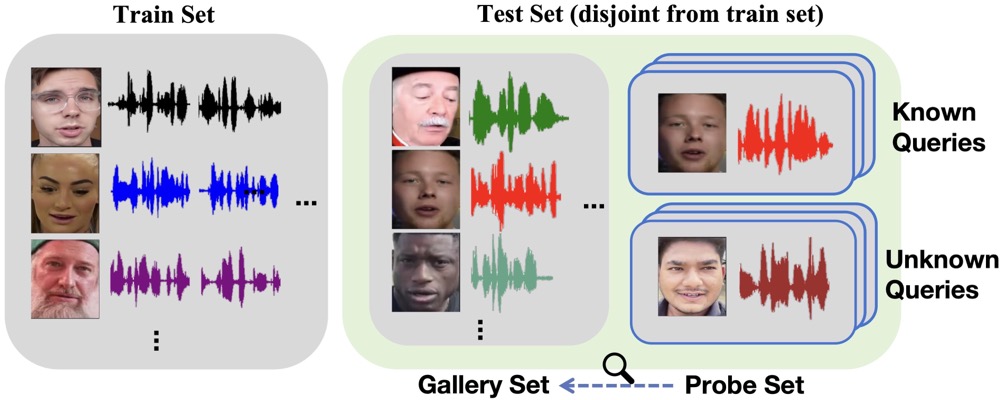}
  \centering
  \caption{{The outline of the Open-Set Speaker-Identification task. The evaluation protocol requires a gallery set for speaker enrollment. Then, the test queries must be linked to the speaker in the gallery as known queries or rejected as unknown queries based on the similarities scores and a pre-defined threshold.  }}
  \label{fig::OSSI}
  \vspace{-2.2em}
\end{figure}
\vspace{-1em}
\section{VoxBlink2 Dataset}
\subsection{Data Description}
The VoxBlink2 corpus is composed of 9,904,382 high-quality utterances and their corresponding video clips, sourced from 111,284 users on YouTube. To our best knowledge, it is the largest publicly available audio-visual dataset for speaker recognition. Unlike the VoxBlink\cite{lin2023voxblink}, which only utilizes short video segments, we extract the initial minute of long-duration user-uploaded videos, significantly expanding the diversity and scenarios of data. Other information is detailed in Tab.\ref{tab::statistics}.
\vspace{-0.5em}
\begin{table}[ht]\centering \footnotesize
\caption{\label{tab::statistics} {\it The statistics for the VoxBlink2 dataset compared with the VoxCeleb2 and the VoxBlink. Utt and Dur mean utterance and duration, respectively. }}
\vspace{-0.5em}
\begin{tabular}{@{}lccc@{}}
\toprule
\textbf{Dataset} & \textbf{VoxCeleb2} & \textbf{VoxBlink} & \textbf{VoxBlink2}  \\ 
\midrule
\# of SPKs  & 5,994 & 38,065 & 111,284 \\
\# of videos & 145,569 & 372,084 & 2,097,062 \\
\# of hours & 2,442 & 2,135 & 16,672 \\ 
\# of utterances & 1,092,009 & 1,455,190&  9,904,382 \\
Avg \# of Utts per spk & 185 & 38.23 & 89 \\
Avg \# of Dur. per Utt & 7.8 & 5.3 & 6.0 \\
Avg \# of Span(days) per spk & - & 440 & 786 \\
\bottomrule
\end{tabular}
\vspace{-2em}
\end{table}
\vspace{-0.5em}
\subsection{Data mining}

\subsubsection{Collection Pipeline}\label{sec::pipeline}
While the data mining process for VoxBlink2 follows a pipeline similar to that of VoxBlink, several modifications have been implemented to enhance data quality further. The stem pipeline can be outlined as follows:

\textbf{Step I. Candidate Collection.} Considering the impact of language diversity on speaker recognition systems, we compile a long keyword list spanning 18 languages for user retrieval. Then, we collect over 6 million 1-min videos from Youtube users who utilize their photos as avatars. It is noteworthy that we intentionally avoid duplicate users with the VoxBlink and duplicate recordings with the VoxCeleb1\&2.

\textbf{Step II. Frame Extraction \& Face Detection.} In pursuit of higher quality and efficiency, we employ a high frame rate (25 fps) for frame extraction and utilize the MobileNet\cite{howard2017mobilenets} to detect facial movements. The 1-person video tracks are generated by setting a threshold of the minimum Intersection Over Union (IOU) value between two consecutive units, ensuring that each facial track includes only one person.

\textbf{Step III. Face Recognition.}  After the face detection, we identify the faces along the video track by our pre-trained ArcFace classifier, which is introduced in \ref{sec::arcface}. Adopting the identification approach rather than verification enhances data purity.

\textbf{Step IV. Active Speaker Detection \& Overlap Speech Detection.} To mitigate the inclusion of silent and overlapped segments, we integrate an audio-visual speaker diarization model\cite{10095802} and an overlap detection model\cite{cheng2023dku} into our pipeline. These models enable the partitioning of active speech segments and eliminating overlapping speech segments, respectively.
\subsubsection{ Face Classifier Training}\label{sec::arcface}

\begin{figure}
  \includegraphics[width=0.34\textwidth]{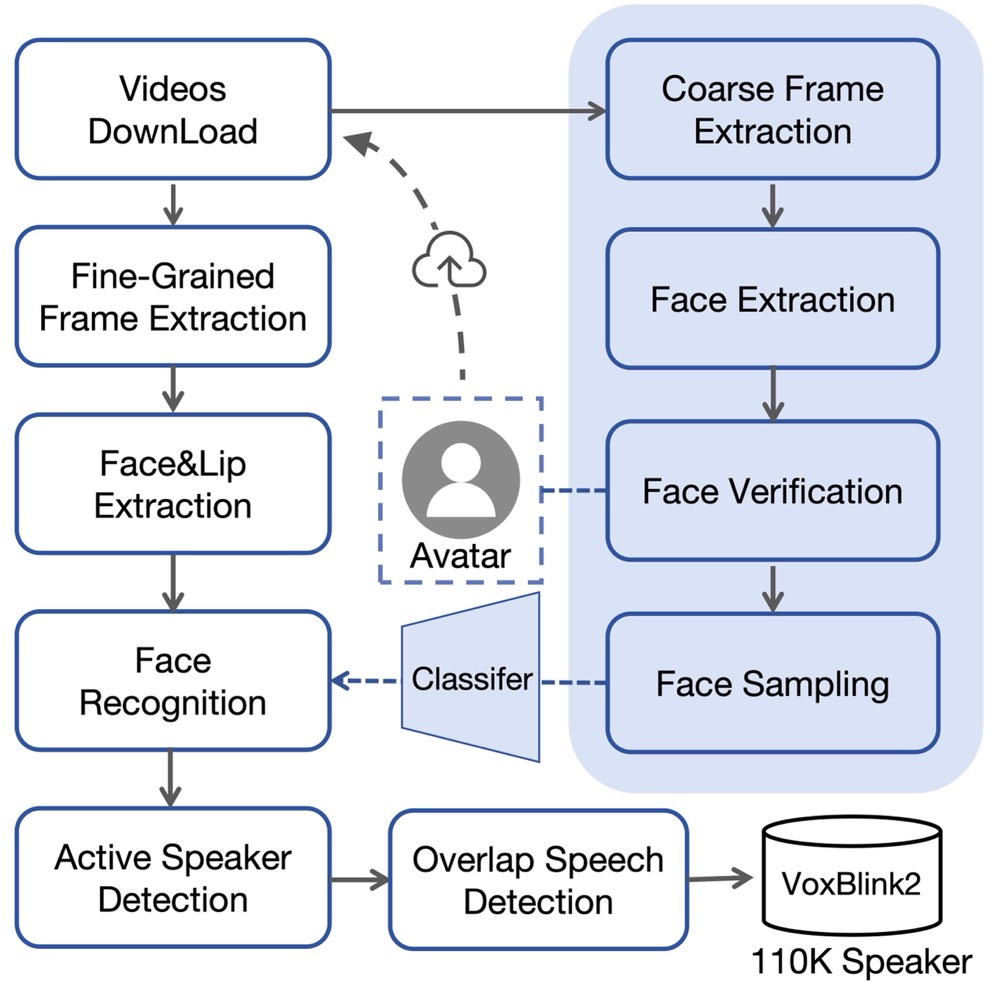}
  \centering
  \caption{{The outline of the data collection pipeline. The modules highlighted in blue(right) are designed to extract faces from candidates for better accuracy.}}
  \label{fig::pipeline}
  \vspace{-2em}
\end{figure}

The accuracy of open-set 1:1 verification is often influenced by inter-domain differences, leading to some error labels in datasets constructed using face verification methods such as VoxBlink. As illustrated in Fig. \ref{fig::pipeline}, we have introduced a \textbf{supplementary branch} (highlighted in blue) dedicated to training a classifier, following the outlined procedures:

\textbf{Coarse Frame Extraction \& Face Detection.} In contrast to Step II described in Sec.\ref{sec::pipeline}, we utilize a relatively low frame rate to capture frames from all candidate videos. Subsequently, frames featuring single-person appearance are exclusively detected by the mobilenet\cite{howard2017mobilenets} for face verification.

\textbf{Face Verification.}  The ResNet-IRSE50 model \cite{8953658} is employed to extract face embeddings from the obtained facial images and compute 1:1 similarity scores with the template embedding from the candidate's avatar photo.

\textbf{Face Sampling}. We use cosine scores to weighted-sample faces from each speaker, with the maximum number of faces capped at 10 per speaker. In short, faces exhibiting higher cosine similarity to the avatar embedding are prioritized for selection for training.

Ultimately, we concatenate ArcFace at the end of the ResNet-IRSE50 encoder and proceed to train the encoder with the classifier by approximately 200K candidate face data collected from face sampling. Furthermore, the VoxBlink2 dataset demonstrates a substantial increase in accuracy, reaching 92\%, compared to the 72\% accuracy achieved by the VoxBlink dataset, as verified through manual assessment of a randomly sampled subset of 50 speakers.

\section{Open-Set Speaker-Identification}
Traditional 1:1 verification tasks are typically deployed on terminal devices, whereas the m: n open-set identification offers an opportunity to integrate speaker recognition systems into practical access control systems. Despite this potential,  the exploration of this topic in speaker recognition remains limited, partly due to the absence of large-scale evaluation protocols. Although \cite{peri2023voxwatch} and \cite{malegaonkar2011performance} have made some endeavors, the open-source evaluation sets are not publicly accessible to academia, and the scale of data is relatively limited. Motivated by the methodology proposed in \cite{jain2011handbook,gunther2017toward} for face recognition, our proposed protocol aims to establish a standardized framework for evaluating speaker recognition systems in open-set identification scenarios, with the ability to recognize enrolled speakers and reject unprivileged speakers.

\subsection{Evaluation Protocols}
As depicted in Fig.\ref{fig::OSSI}, the evaluation protocols involve two different sets: gallery set ($\textbf{\textit{G}}$) and probe set ($\textbf{\textit{P}}$). The gallery set ($\textbf{\textit{G}}$) is constructed with an equal number of enrollment utterances for each known identity, essentially serving as a knowledge base. On the other hand, $\textbf{\textit{P}}$ is dedicated to querying $\textbf{\textit{G}}$ and is divided into two categories: Known Queries ($\textbf{\textit{K}}$) and Unknown Queries ($\textbf{\textit{U}}$). Here, we adopt the VoxBlink-clean set (1,028,095 utterances from 18,381 speakers) and fabricate the following protocols in terms of different application scenarios, which is shown in Tab.\ref{tab::VB-Eval}. Furthermore, we set for each protocol with 1, 3, and 5 enrollment utterances per speaker in $\textbf{\textit{G}}$, therefore adding up to 3*3=9 different evaluation protocols.
\begin{table}[t]\centering \footnotesize
\caption{\label{tab::VB-Eval} {\it Basic information of three evaluation protocols. The numbers denote the numbers of speakers in gallery, known and unknown query sets. S, M, L indicate Small, Medium, Large. }}
\vspace{-0.5em}
\begin{tabular}{@{}lcccc@{}}
\toprule
~ & $\textbf{G}$ & $\textbf{K}$ & $\textbf{U}$ & \textbf{Authentication Scenarios} \\ 
\midrule
\textbf{VB-Eval-S}  & 60 & 30 & 30 & Exam room\\
\textbf{VB-Eval-M}  & 600 & 300 & 300 & Office building  \\
\textbf{VB-Eval-L}  & 6,000 & 3,000 & 3,000 & Major events \\

\bottomrule
\end{tabular}
\vspace{-2em}
\end{table}
\vspace{-0.5em}

\subsection{Evaluation Metrics}
The OSSI can be measured by the Detection and Identification Rate (DIR) and False Alarm Rate (FAR)\cite{jain2011handbook}. Moreover, we adopt the DIR@FAR to recognize known speakers while maintaining a fixed FAR threshold. 

Initially, upon obtaining a feature extractor trained on the dataset, speaker embeddings are simultaneously extracted from $\textbf{\textit{G}}$ and $\textbf{\textit{P}}$, where ($\textbf{\textit{P}}=\textbf{\textit{K}}\cup\textbf{\textit{U}}$). When a known probe $p$ is presented to a system, the similarity scores between $p$ and all samples in $\textbf{\textit{G}}$ are computed and sorted. A probe $p$ has rank $n$ if $s(p,G_p)$ is the n-th largest speaker embedding similarity score, here $G_p$ represents the matched speaker with $p$ in $\textbf{\textit{G}}$.

Finally, for a given pre-defined similarity threshold $\theta$ and a rank $n$, the DIR can be derived by Eq.\ref{con::DIR}. In our evaluation framework, the n is set to 1 to calculate the top-match speaker, and we use the cosine-similarity for similarity calculation. 
\vspace{-0.5em}
{
\begin{equation}
\begin{aligned}
DIR(\theta,n) = \frac{\lvert{ rank(p)\geq n \wedge sim(p,G_p)\geq \theta;p\in \textbf{\textit{K}}}\rvert}{\lvert \textbf{\textit{K}} \rvert}
\label{con::DIR}
\end{aligned}
\end{equation}
}
The False Alarm Rate (FAR) serves as a measure of the system's ability to discern and reject unknown queries, typically considered impostors. A false alarm event occurs when the top match score for an imposter in $\textbf{\textit{U}}$ is higher than $\theta$. Assuming that $G_p$ symbolises the most-matched speaker with $p$ in $\textbf{\textit{G}}$, the FAR can be computed by:
\vspace{-0.25em}
{
\begin{equation}
\begin{aligned}
FAR(\theta) = \frac{\lvert{ sim(p,G_p)\geq \theta;p\in \textbf{\textit{U}}}\rvert}{\lvert \textbf{\textit{U}} \rvert}
\label{con:FAR}
\end{aligned}
\end{equation}
}
The optimal system should have a 1.0 DIR and a 0.0 FAR, indicating perfect detection and identification of all individuals in the probe set without any false alarms. However, in real-world systems, there is a trade-off between the DIR and the FAR. This trade-off is influenced by varying threshold values ($\theta$) that can be adjusted to meet specific FAR requirements, and visualizing this trade-off is often done through a Receiver Operating Characteristic (ROC) curve. For some well-defined authentication scenarios with pre-defined FAR, we can use the DIR at a particular FAR (DIR@FAR) value as a metric.

\section{Experimental Settings}
\textbf{Data Usage.}  Our experiments primarily utilize the VoxBlink2 (VB2) and VoxCeleb2 (VC2) datasets for training, while evaluations are conducted on the VoxCeleb1 test set for the speaker recognition task and the VB-Eval dataset for the OSSI task. Moreover, we compile several subsets of VB2 to investigate the influence of data scale on model performance. The acoustic features are 80-dimensional log Mel-filterbank energies with a frame length of 25ms and a hop size of 10ms. The input frame length is fixed at 200 or 500 frames. 

\textbf{Model Usage.} Our approach employs ResNet-based\cite{cai18_odyssey} models of various sizes, complemented by two pooling methods: Attentive Statistic Pooling (ASP)\cite{okabe18_interspeech} and Temporal Statistic Pooling (TSP). To further harness the latent potential of data, we employ the Simple Attention Modules (SimAM) to extract more discriminative speaker embeddings\cite{yang2021simam,qin2022simple}. Additionally, we introduce the ResNet50-based face recognition model for comparative analysis. This model is trained by Glint360K\cite{an2021partial}, which comprises 17,091,657 faces of 360,232 individuals.

\textbf{Training strategies.} We incorporate two different strategies for model training as follows:
\begin{itemize}

\item \textbf{Pre-train on the VC2 and fine-tune on the mixed set.} Following the findings in the \cite{farfield_xiaoyi}, the Mix-FT strategy demonstrates the capability to further enhance the performance of speaker recognition systems.
\item \textbf{Pre-train on the VB2 and fine-tune on the VC2 set.} Inspired by the findings of the LLM, models trained with massive data exhibit stronger generalization abilities. Moreover, by fine-tuning the VC2 set, the highly generalized model can learn more refined features.
\end{itemize}
Specifically, in both pre-training stages, we adopt the on-the-fly data augmentation for the variation of data and the speed perturbation to triple the number of speakers. The SGD optimizer updates the model parameters, and the StepLR scheduler with an initial learning rate of 0.1 decays to 1e-4 until convergence. For the fine-tuning stage, the Large-Margin Fine-Tune (LMFT) strategy\cite{9414600} is introduced, accompanied by the removal of data augmentation. The LR in this phase must be set lower than the pre-training phase, and employing a relative smaller LR for a larger model has been found to be more effective.

\section{Results}

\subsection{Speaker Verification}

\subsubsection{Different Strategies}
We adopt the ResNet100-ASP with the simple attention module as the speaker encoder to generate speaker embeddings. In addition, we curate several randomly sampled subsets of the VB2 dataset, each containing varying speaker counts: 5k, 10k, 30k, and the full version with 110k speakers. As shown in Tab.\ref{tab::Strategy}, increasing the number of mixed fine-tuned speakers does not consistently lead to significant performance improvements. Besides, compared to the Mix-FT mentioned previously, Fine-tuning the model on a large-scale pre-trained dataset results in a notable 43.4\% in EER reduction on the VoxCeleb1-O set. Since this strategy is more intuitive and effective, the following experiments follow this training pattern.

\begin{table}[b]\centering \footnotesize
\vspace{-2em}
\caption{\label{tab::Strategy} {\it The system performance on the VoxCeleb1-O test set based on different training data for different stages. The LMFT and other post-processing strategies have not been introduced. }}
\begin{tabular}{@{}llcc@{}}
\toprule
\textbf{ Pre-train }&\textbf{Fine-tune}& \textbf{EER[\%]} & \textbf{$\textbf{minDCF}_{0.01}$} \\ 
\midrule
VC2  & - & 0.606 & 0.052 \\
VC2  & VC2+VB2-5K & 0.527 & 0.047 \\
VC2  & VC2+VB2-10K & 0.505 & 0.049  \\
VC2  & VC2+VB2-30K & 0.505 & 0.051 \\
VC2  & VC2+VB2 & 0.674 & 0.066 \\
\midrule
VB2  & - & 0.893 & 0.093 \\
VB2  & VC2 & \textbf{0.340} & \textbf{0.026} \\
\bottomrule
\end{tabular}
\end{table}
\vspace{-1em}
\subsubsection{Different Pre-train Data Scales}
\begin{figure}[t]
  \includegraphics[clip, trim=0cm 12cm 0cm 0cm,width=0.44\textwidth]{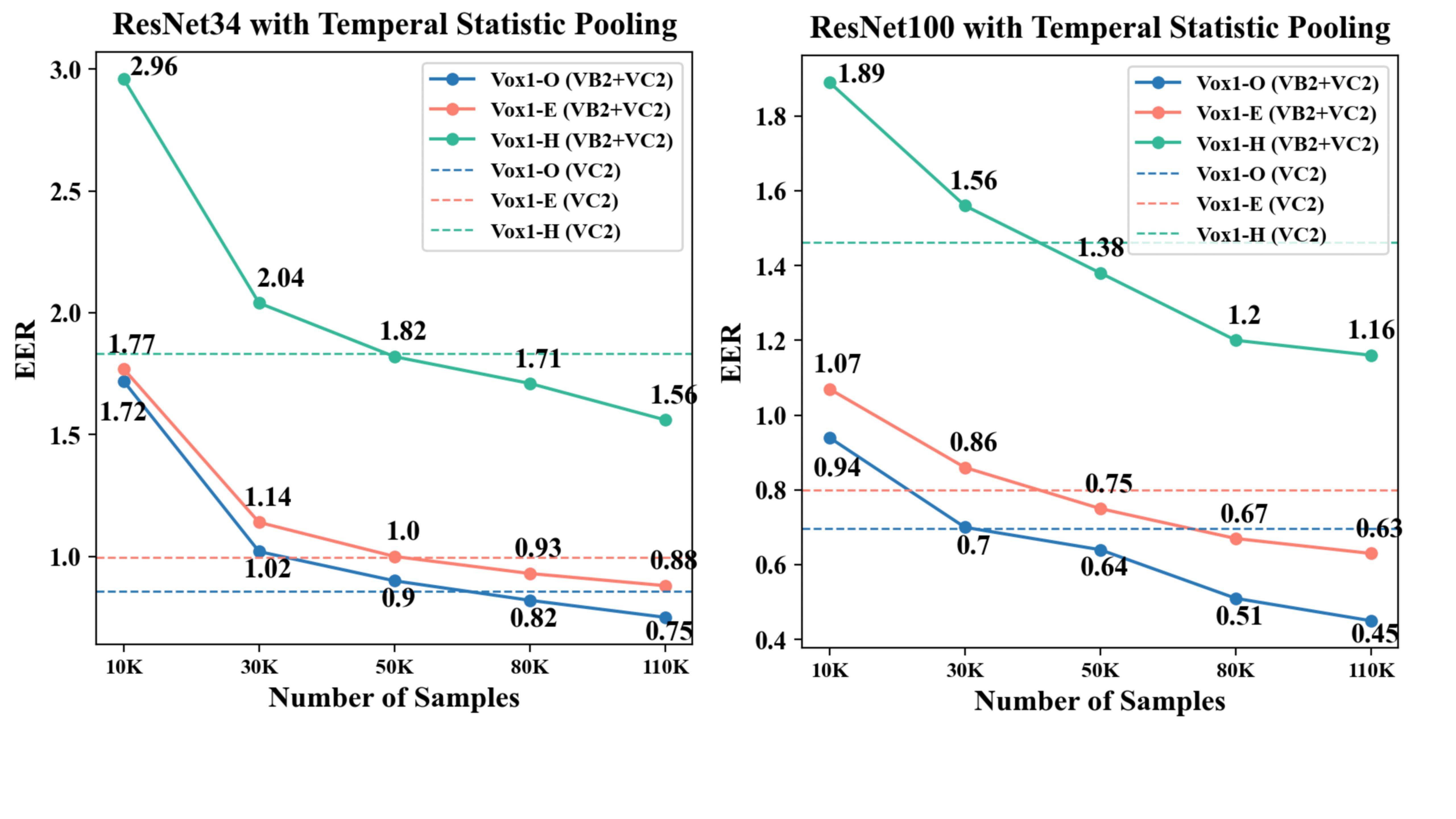}
  \vspace{-0.5em}
  \caption{{The EER and minDCF performance on VoxCeleb1-test set. ResNet34-TSP and ResNet100-TSP models, trained with different data scales, are pre-trained and then fine-tuned on VC2. Dotted lines represents system performances directly trained on VC2 for comparison. LMFT and other post-processing strategies are not included in this analysis.}}
  \vspace{-1em}
  \label{fig::DataScale}
\end{figure}

For a more detailed examination of performance variations with changes in data volume, we randomly compile diverse sub-sets of the VB2 set with different size of speakers. As illustrated in Figure \ref{fig::DataScale}, an increase in the number of speakers correlates with enhanced performance. It can be indicated that by stacking more data during the pre-training phase, the model becomes more robust and generalized to adapt to diverse domains. From another perspective, in line with the scaling law principle, amplifying the data volume requires enlarging the model to achieve more significant effects, which is also illustrated in Fig.\ref{fig::DataScale}. When we increase the number of model parameters, the decreases in EER on all test sets become steeper.

\begin{figure}[t]
  \includegraphics[clip, trim=0cm 10cm 2cm 1cm,width=0.44\textwidth]{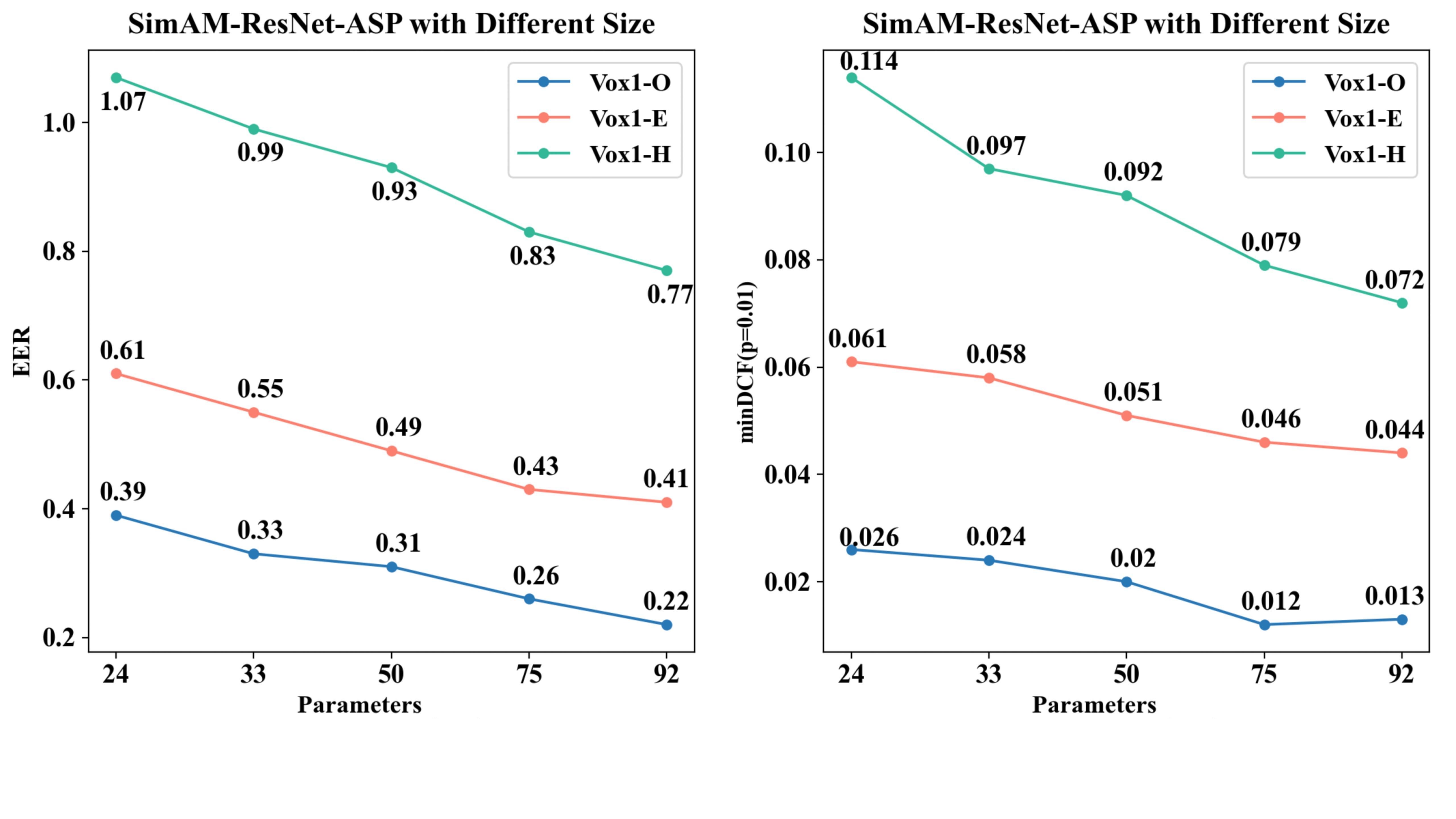}
  \vspace{-1em}
  \caption{{The EER and minDCF performances on VoxCeleb1-test set. All the ResNet-based models (ResNet{34, 50, 100, 221, 293}) embedded with the SimAM and the ASP, are pre-trained on the full VB2 set and then fine-tuned on VC2. The LMFT strategy is introduced during the fine-tuning stage. }}
  \label{fig::ModelScale}
  \vspace{-2.5em}

\end{figure}

\begin{table}[b]\centering \footnotesize
    \vspace{-2em}

    \caption{\label{tab::Post-Processing} {\it The post-processing results based on the SimAM-ResNet293 single system. The $p_{target}$ is set 0.01.}}
    \vspace{-0.5em}
    \begin{tabular}{@{}lcccccc@{}}
    \toprule
    \multirow{2}*{\textbf{Method}} & \multicolumn{2}{c}{\textbf{Vox1-O}} & \multicolumn{2}{c}{\textbf{Vox1-E}} & \multicolumn{2}{c}{\textbf{Vox1-H}}\\
    \cmidrule(lr){2-7}& \textbf{EER} & \textbf{mDCF} & \textbf{EER} & \textbf{mDCF}& \textbf{EER} & \textbf{mDCF}  \\
    \midrule

    ResNet293  & 0.23 & 0.013 & 0.42 & 0.044 & 0.77 & 0.072 \\
    +AS-Norm  & 0.22 & 0.009 & 0.40 & 0.042 & 0.73 & 0.073 \\
    ++QMF  & \textbf{0.17} & \textbf{0.006} & \textbf{0.37} & \textbf{0.037} & \textbf{0.68} & \textbf{0.070} \\
    \bottomrule
    \end{tabular}
\end{table}

\vspace{-0.5em}
\subsubsection{Different Model Complexity}
To explore the performance bounds brought about by the increase in data volumes, we progressively escalate the model complexity. As shown in Fig.\ref{fig::ModelScale}, we observe continuous boosting performance on the Vox1-test set with the model size expansion.

Furthermore, we adopt the same settings as \cite{li2023dku} for post-processing on the ResNet293-based model and finally achieve state-of-the-art performance. As shown in Tab.\ref{tab::Post-Processing}, the EER and the minDCF can be reduced to 0.17\% and 0.006\% on the VoxCeleb1-O test set, respectively.
\vspace{-0.5em}
\subsection{Open-Set Speaker-Identification}
\vspace{-0.5em}
To assess the influence of data scale and modality on the OSSI, we adopt the ResNet50 as the backbone to train the feature extractor and utilize the DIR at different FARs to evaluate the OSSI performance. As depicted in Tab.\ref{tab::Result_OSSI}, increasing the number of enrollment utterances exhibits a positive association with the enhancement of DIR@FAR. However, as the gallery size expands exponentially, a noticeable decline in performance is observed, indicating the need for further studies.

\begin{table}[t]\centering \footnotesize
\caption{\label{tab::Result_OSSI} {\it The baseline of the OSSI based on ResNet50 (Pre-trained on VB2, holding a 1.02\% EER on VoxCeleb-O test set), reflecting the DIR performance at different FAR. Enroll nums means the number of utterances included in the gallery set per speaker. The DIR@FAR=1 denotes there is no rejection. }}
\vspace{-0.7em}
\setlength{\tabcolsep}{2.4mm}{
\renewcommand{\arraystretch}{0.9} 
\begin{tabular}{@{}cccccc@{}}
\toprule
\multirow{2}*{\textbf{Protocol Type}} & \multirow{2}*{\textbf{Enroll nums}} & \multicolumn{4}{c}{\textbf{DIR@FAR [\%]}} \\
\cmidrule(lr){3-6} & ~ &  \textbf{0.001} & \textbf{0.01} & \textbf{0.1} & \textbf{1} \\
\midrule

\multirow{3}*{\textbf{VB-Eval-S}} 
& 1  & 88.10 & 93.35 & 93.46 & 98.08   \\
&3 & 96.32 & 97.98 & 98.74 & 99.02\\
&5 & 96.97 & 98.04 & 98.68 & 99.11 \\
\midrule
\multirow{3}*{\textbf{VB-Eval-M}} 
& 1  & 68.80 & 82.95 & 90.40 & 94.36   \\
&3 & 86.86 & 92.37 & 95.95 & 97.69\\
&5 & 91.60 & 94.90 & 96.91 & 98.22 \\
\midrule

\multirow{3}*{\textbf{VB-Eval-L}} 
& 1  & 20.73 & 66.37 & 80.95 & 88.09   \\
&3 & 23.17 & 83.65 & 92.66 & 95.86\\
&5 & 24.94 & 87.12 & 94.34 & 96.72 \\

\bottomrule

\end{tabular}
}
\end{table}

\begin{figure}[t]
\vspace{-1em}

  \includegraphics[clip, trim=0cm 38cm 0cm 0cm,width=0.47\textwidth]{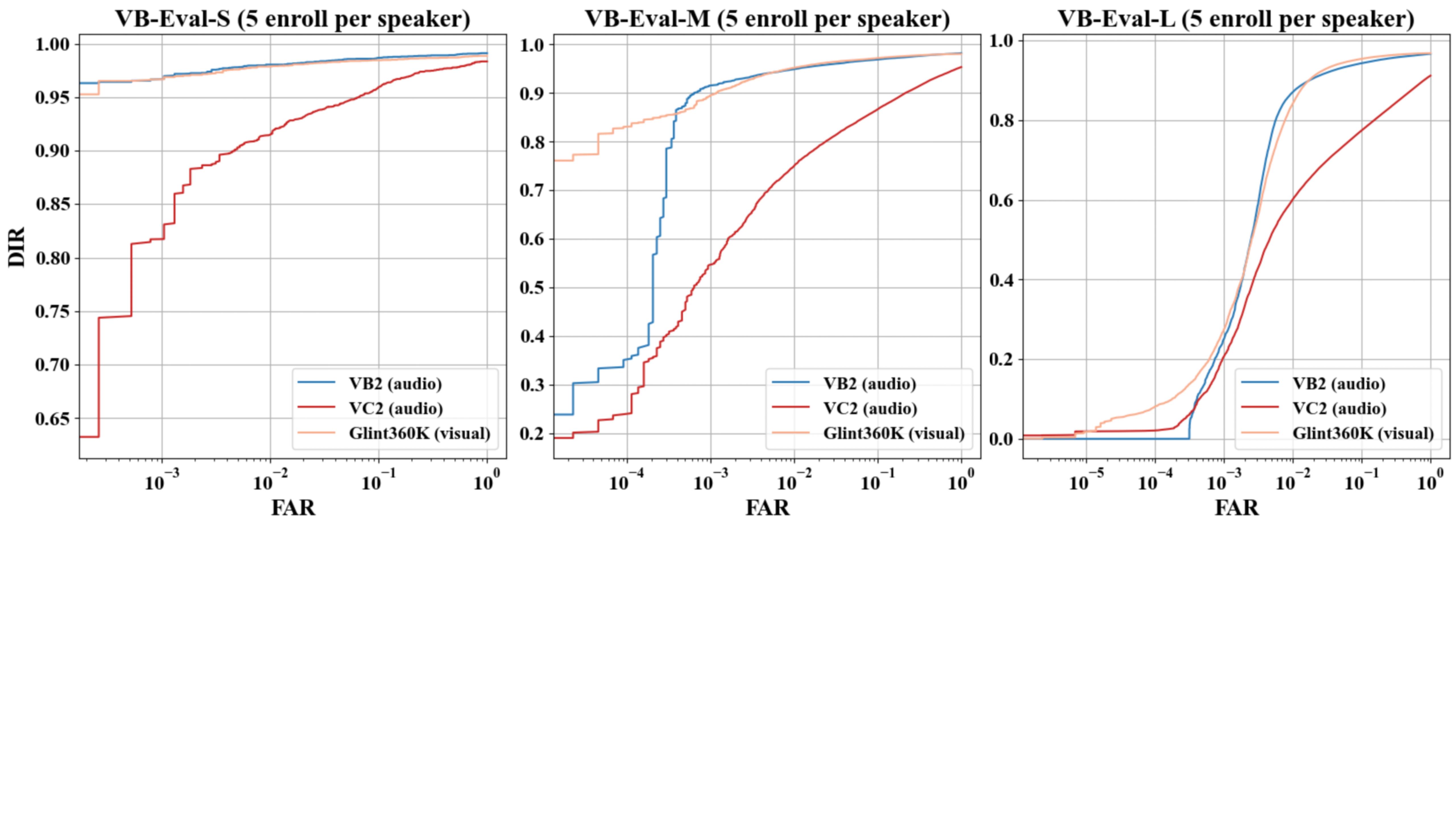}
  \caption{{ The ROC curve of two speaker recognition models and a face recognition model. }}
  \vspace{-0.5em}
  \label{fig::ROC}
\vspace{-2em}

\end{figure}

 In another modality, we utilize the same backbone pre-trained with Glint360K as the face recognition model, achieving a remarkable 0.03\% EER on the VoxCeleb1-O test set\cite{tao2023speaker}. Although the training dataset of speaker recognition differs from that of face recognition, the scales of data are comparable (110K vs 360K). Subsequently, we extract faces from videos in both the gallery and probe sets at intervals of 0.3 seconds. By encoding the faces from the same video and then averaging the face embeddings, we obtain the utterance-level face embedding.
 
 As shown in Fig.\ref{fig::ROC}, we observe that large-scale training of facial and speaker recognition models yields relatively comparable results in both VB-Eval-S and VB-Eval-L. While the models show outstanding results in the VB-Eval-S, they reflect degraded performance in the VB-Eval-L. We speculate that the key factor is the insufficient training data for both modalities. Additionally, there is a slight discrepancy in performance between the speaker and face models in VB-Eval-M. However, the distance between the speaker model and the face model for recognition can be narrowed by enlarging the quantity of data.
 
\section{Conclusion}
This paper provides a large-scale audio-visual corpus for speaker recognition, comprising over 110K individuals gathered from YouTube. Through a series of ablation studies, we investigate the impact of training strategies, data scale, and model complexity on speaker verification, achieving state-of-the-art results. In addition, we introduce a new Open-Set Speaker-Identification benchmark alongside relevant baseline metrics derived from the VoxBlink-clean dataset. Notably, our findings reveal that speaker recognition models trained on comparable data scales and utilizing similar architectures as facial recognition models demonstrate comparable performance.

\section{Acknowledgement}
This research is funded by China Mobile Research Institute-Wuhan University Systematic Artificial Intelligence Collaboration Platform under the project "Joint Research and Development of Large Voiceprint Model for Telephone Scenarios". Many thanks for the computational resource provided by the Advanced Computing East China Sub-Center.
\bibliographystyle{IEEEtran}
\bibliography{mybib}

\end{document}